\documentclass{Rinton-P9x6}
\begin{document}

\title{Properties of metallic films in precise calculation of the Casimir force}
\author{Vitaly B. Svetovoy}

\address{Transducers Science and Technology Group, EWI,
University of Twente, \\P.O. 217, 7500 AE Enschede, The
Netherlands
\\E-mail: v.b.svetovoy@el.utwente.nl
}
\maketitle

\abstracts{ Optical properties of the deposited gold films are
discussed in connection with the Casimir force prediction. Voids
in the films and electron scattering on the grain boundaries
reduce the force at small separations on the $2\;\%$ level in
comparison with the bulk material prediction. The contribution of
the patch potential due to polycrystalline structure of the films
is shown to be small for the existing Casimir force experiments.}

\section{\protect\bigskip Introduction}

Detailed investigation of the Casimir force\cite{Cas} between two
uncharged metallic plates, predicted in 1948, became possible only
recently with the progress in microtechnologies allowing to
control separation between bodies smaller than $1\ \mu m$. The
development of the experiment\cite{Lam,Moh,Allexp,Decca} is
characterized by steady improvement in the precision of the force
measurement from $5\;\%$\cite{Lam} to $0.25\;\%$\cite{Decca}. The
progress in theoretical prediction of the force is not so
impressive. Starting from the first precise
calculations\cite{BS,LR00,KMM,SL2} and up to now\cite{Decca} the
force is calculated using the optical properties of bulk metals
taken in the handbooks. However, in all the experiments the force
was measured between deposited metallic films. Typically it was
gold with the thickness $100-200\;nm$. It is clear that the
deposited material and bulk material are different if we pretend
on high precision calculations. In this paper the question is
analyzed to what extend the properties of real deposited metallic
films can change the force in comparison with that predicted for
the bulk material. To answer this question we inevitably have to
turn to the material science. The steps in this direction never
been done before but without understanding of the used material
there will be no further progress in the prediction of the Casimir
force.

Originally the Casimir force\cite{Cas} was calculated between
ideal metallic plates. In this case it does not include any
material parameters: $F_{c}=\left( \pi ^{2}/240\right) \left(
\hbar c/a^{4}\right) $, where $a$ is the separation between
plates. At small separations the deviation of the used metal from
the ideal one becomes significant and one has to calculate the
force using a more general Lifshitz formula\cite{LP9}. This
formula takes into account real optical properties of the plate
material via the dielectric function $\varepsilon \left(
\omega\right) $, which is taken in the handbooks\cite{HB1}. Gold
is the best material for the force measurement since it is
chemically inactive and its low frequency behavior, where the
handbook data unaccessible, can be reliably predicted with the
Drude model.

\section{Voids in the films}

Optical properties of $Au$ films were investigated widely in 60$%
^{th}$-70$^{th}$. The results are collected in the
handbooks\cite{HB1}. Significant deviation in \ the optical data
was attributed to genuine sample differences caused by different
sample preparation methods\cite{HB1}. The handbook data\cite{HB1}
were carefully chosen to represent the bulk material as close as
possible. A special investigation of the sample preparation effect
was undertaken\cite{AKB} with the conclusion that the most
significant role played voids in the samples. Annealed films
demonstrated larger density due to larger grain size. A single
parameter model representing voids in an otherwise homogeneous
medium was shown to account for the major discrepancies in the
above-band-gap ($E=2.5\;eV$) spectra for $Au$ samples prepared in
different ways. It allowed to get information about the volume
fraction of voids $f_{V}$ in different samples. For $f_{V}\ll 1$
and $\left| \varepsilon \right| \gg 1$ the model gave the
effective dielectric function $\left\langle \varepsilon
\right\rangle $ defined by the relation $\left\langle \varepsilon
\right\rangle =\varepsilon \left( 1-\frac{3}{2}f_{V}\right) $. For
the film ($150\;nm$) evaporated with
e-beam on $NaCl$ substrate at room temperature with a deposition rate of $%
23\;$\AA $/s$ the volume of voids was found to be larger on
$4\;\%$ than in the annealed samples used for the handbook
data\cite{HB1}. For the $50\;nm$ thick sputtered films deposited
at the rate smaller than $13\;\AA/s$ the volume of voids was found
to be $4\;\%$ at room temperature deposition and $1\;\%$ at
$250^{\circ }\;C$. The cleaved $NaCl$ substrate was chosen because
the films start to grow epitaxially on it. The same conclusion was
made in a more recent investigation\cite{Wang}.

In the most sensitive experiments the $Au$ films were
evaporated\cite{Moh} or sputtered\cite{Decca} at room temperature,
no annealing was reported. One can expect\cite{AKB} that in both
cases the volume of voids was of about of $4\;\%$.

The correction to the Casimir force due presence of voids was
calculated using the data\cite{HB1} for $\varepsilon $. These data
were scaled in accordance with the effective dielectric function
$\left\langle \varepsilon \right\rangle =\varepsilon \left(
1-\frac{3}{2}f_{V}\right) $. Calculations were performed in the
way similar to \cite{LR00} for plate-plate and sphere-plate
geometries. The results for the relative correction to the force
are shown in Fig. \ref{fig1}(\textit{a}) for minimal separations
in the experiments: $a=63\;nm$ for sphere-plate\cite {Moh} and
$a=260\;nm$ for the effective plate-plate geometries\cite{Decca}.
Let us stress that this correction always make the force for films
smaller than that for the bulk material.

\section{Scattering on the grain boundaries}

It is well known that the resistivity of deposited films deviates
significantly from the bulk resistivity. If the film is not very
thin, say thicker than $10\;nm$, it has polycrystalline structure.
The grain boundaries contribute to the electron transport (for the
references see\cite{SEN}). In a recent paper\cite{SEN} the optical
characteristics of the films were treated as the grains with bulk
material properties plus electron scattering on the grain
boundaries.

The gold films with the grain size from 15 to 45 $nm$ were
deposited on glass with two different methods\cite{SEN}.
Reflectance of the films was investigated in the wavelength range
0.3-50 $\mu m$. The grain boundaries were modelled with the delta
potentials, scattering from phonons and point defects were
accounted by the relaxation time. The grain size distribution had
no significant influence, so the films were characterized only by
the mean grain size $D$. The material is described by the
two dielectric functions longitudinal $\varepsilon _{l}$ and transverse $%
\varepsilon _{t}$. Only $ \varepsilon _{t}$ changes because the
grain boundaries are perpendicular to the film surface. In the
local limit for $\varepsilon _{t}$ the following expression was
found

\begin{eqnarray}
\varepsilon _{t}\left( \omega \right)  &=&\varepsilon \left( \omega \right) +%
\frac{3\omega _{p}^{2}}{\omega \left( \omega +i\omega _{\tau }\right) }\left[
\frac{1}{2}\alpha -\alpha ^{2}+\alpha ^{3}\ln \left( 1+\alpha ^{-1}\right) %
\right] ,\quad   \label{epst} \\
\alpha  &\equiv &\frac{v_{F}}{D\omega _{\tau }}\frac{\mathcal{R}}{1-\mathcal{%
R}}\left( 1-i\frac{\omega }{\omega _{\tau }}\right) ^{-1},
\end{eqnarray}

\noindent where $\varepsilon \left( \omega \right) =\varepsilon
_{l}\left( \omega \right) $ is the dielectric function of the
crystalline gold, $\omega_{\tau}$ is the Drude relaxation
frequency, $v_{F}$ is the Fermi velocity, $\mathcal{R}$ is the
reflection coefficient of electrons on the grain boundary. The
latter is the
only empirical parameter, which was found in the same experiment to be $%
\mathcal{R}\approx 0.65$. The reflectance of the film was
expressed via the surface impedance. Comparison with the
experimental reflectance showed that the model describes well the
effect of the grain size except for the very small grains
$D\approx 15\;nm$ when the electron mean free path becomes
comparable with $D$.

\begin{figure}[t]
\epsfxsize=11.8cm 
\epsfbox{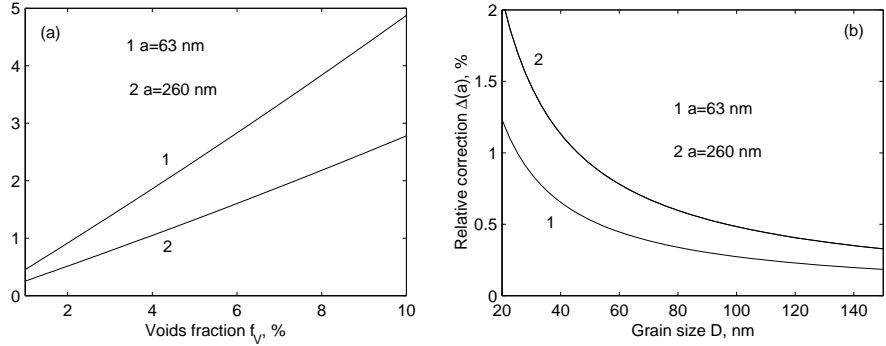} 
\caption{The relative corrections to the Casimir force $\Delta(a)$
in percents for the minimal separations explored in the
experiments $a=63\;nm$ \protect\cite{Moh} and $a=260\;nm$
\protect\cite{Decca}: (a) correction due to voids in the film; (b)
correction due to scattering on the grain boundaries as a function
of the mean grain size $D$} \label{fig1}
\end{figure}

We used the expression (\ref{epst}) at imaginary frequencies
$\omega =i\zeta $ and calculated the impedances for $s$ and $p$
polarizations following the Kliewer and Fuchs procedure\cite{KF}.
The impedances are different due to the nonzero wave vector
$\mathbf{q}$ along the plates in the Lifshitz formula. The details
will be reported later. The relative correction to the Casimir
force as a function of the grain size is presented
in Fig. \ref{fig1}(\textit{b}). The typical grain size for the films $%
100-200\;nm$ thick deposited on the substrate at room temperature
is less than $50\;nm$\cite{VSHOVA} but it can depend on the
deposition details and we take as the upper limit $D<100\;nm$.

\section{Patch potential}

The importance of the patch potential for the Casimir force
measurement was stressed in a recent paper\cite{ST}. For a crystal
the work function has different values for different
crystallographic planes. This difference should be compensated by
the potential distribution around the crystal. The deposited films
are polycrystalline and for this reason there will be local
variation of the potential nearby the surface. It will result in
an additional force, electrostatic in nature, which will be
measured together with the Casimir force. A simple model was
proposed\cite{ST} to estimate the effect. In this model the
dipoles were distributed on the planes. Their interaction gives
the force which is defined by the spectral density of the surface
potential correlation function $C(k)$. The model is in agreement
with the independent \textit{ab initio} calculation of the
potential distribution near $Al$ crystallite\cite{FBB}. For the
sphere and plate the force was written in the form

\begin{equation}
F_{patch}\left( a\right) =-\varepsilon _{0}R\int\limits_{0}^{\infty }dkk^{2}%
\frac{C\left( k\right) e^{-ka}}{\sinh ka},  \label{patch}
\end{equation}

\begin{equation}
C\left( k\right) =\int d^{2}\mathbf{r}e^{-i\mathbf{kr}}c\left( \mathbf{r}%
\right) ,\quad c\left( \mathbf{r}\right) =\left\langle v\left( \mathbf{r}%
\right) v\left( 0\right) \right\rangle ,  \label{spectr}
\end{equation}

\noindent where $\varepsilon _{0}$ is the permittivity of vacuum,
$R$ is the sphere radius, $v\left( \mathbf{r}\right) $ is the
local potential on one of the surfaces and the potential
distribution on different surfaces was assumed to be uncorrelated.
The spectral density $C\left( k\right) $ is crucial for the force
estimate. To our opinion, in the original paper\cite{ST} it was
overestimated. There was no clear idea around which value of $k$
the spectral density should be centered and what is the width of
this function, so these values were chosen quite arbitrary.
Looking at the films surface images one can easy guess that the
spectrum should be centered around the mean grain size $D$ with
the wavelengths presented roughly from $D/2$ to $2D$. If we
consider the spectrum in this range as flat, the force will be the
following

\begin{equation}
F_{patch}\left( a\right) \approx -\frac{4\pi \varepsilon _{0}\sigma _{v}^{2}%
R}{15a}\exp \left( -\frac{2\pi a}{D}\right) .  \label{patchest}
\end{equation}

\noindent Here $\sigma _{v}^{2}$ is the variance of the potential
distribution that was estimated\cite{ST} as $\left( 90\;mV\right)
^{2}$. The relative contribution of the patch potential to the
measured Casimir force is estimated to be smaller than $0.4\;\%$
for the minimal separation $a=63\;nm$ in the AFM
experiment\cite{Moh} and the largest grain size $D=100\;nm$. It
decreases fast for smaller grain size or for larger separations.
However, it should be stressed that the patch force mimics the
Yukawa interaction and can be easily confused with a new force.

\section{Discussion}

Above it was demonstrated that the optical properties of deposited
gold films differ from those of the bulk material. The main
reasons for the deviations are the voids in the films and electron
scattering on the grain boundaries. Both effects diminish the
absolute value of the predicted Casimir force on the one percent
level. We define the relative correction as

\begin{equation}
\Delta \left( a\right) \equiv \left| \frac{F_{f}\left( a\right) -F_{b}\left(
a\right) }{F_{b}\left( a\right) }\right| ,  \label{relcor}
\end{equation}

\noindent where the indexes $b$ and $f$ refer to the bulk and film
materials, respectively. For the AFM experiment\cite{Moh} in the
sphere-plate geometry the expected correction was found for the
most probable values $f_V=4\,\%$ and $D=50\;nm$: $\Delta \left(
63\;nm\right) =\left( 1.9+0.5\right) \,\%=2.4\,\%$. The same
correction for the MEMS experiment\cite{Decca} with the effective
plate-plate geometry is $\Delta \left( 260\;nm\right) =\left(
1.0+0.9\right) \,\%=1.9\,\%$. One can see that these corrections
cannot be ignored since in both experiments the precision was
better than $1\;\%$.

There is an additional effect which also reduces the absolute
value of the predicted force. At frequencies larger than $\omega
_{p}$ the charge density fluctuation (plasmons) can propagate in
the material decreasing the reflection coefficient for
$p$-polarization. Importance of this nonlocal effect for the
Casimir force at separations of the order or smaller than plasma
wavelength $\lambda _{p}$ was stressed for the first time in the
paper\cite{Esquiv} where the correction due to plasmon excitation
was found. For the AFM experiment\cite{Moh} it is estimated as $3\;\%$ at $%
a=63\;nm$. Due to larger separation this correction is smaller for
the MEMS experiment\cite{Decca}, where it is $1\;\%$ at
$a=260\;nm$. The plasmon correction was dismissed in Ref.
\cite{Decca} on the basis that the separation was much larger than
the penetration depth $\delta =\lambda _{p}/2\pi $. In this
connection we should note that the separation $a$ has to be
compared with $\lambda _{p}$ but not $\delta $. Otherwise on the
same basis one could conclude that the finite conductivity
correction is negligible at $a\sim \lambda _{p}$ while in reality
it is of about of 50\% (see Fig. 1 in the paper\cite{LR00} where
the transition point at $a\sim \lambda _{p}$ not at $\lambda
_{p}/2\pi $ is clearly seen).

\textit{In conclusion}, we demonstrated that it is important to
take into consideration the optical characteristics of the
metallic films used in the experiments and showed the way how the
deviation from the bulk metal properties can be estimated. Voids
in the films, scattering on the grain boundaries so as nonlocal
effects all of them tend to reduce the force in comparison with
the bulk metal prediction. Even without nonlocal effects the
reduction is expected on the level of $2\;\%$.

\section*{Acknowledgments}

This work was supported by the Dutch Technology Foundation


\begin{thebibliography}{99}
\bibitem{Cas}  H. B. G. Casimir, Proc. K. Ned. Akad. Wet. \textbf{51}, 793
(1948).

\bibitem{Lam}  S. K. Lamoreaux, Phys. Rev. Lett. \textbf{78}, 5 (1997);
\textbf{81}, 5475 (1998).

\bibitem{Moh}  B. W. Harris, F. Chen, and U. Mohideen, Phys. Rev. A \textbf{62%
}, 052109 (2000).

\bibitem{Allexp}  T. Ederth, Phys. Rev. A \textbf{62}, 062104 (2000); H. B.
Chan, V. A. Aksyuk, R. N. Kleiman, D. J. Bishop, and F. Capasso,
Science \textbf{291}, 1941 (2001); G. Bressi, G. Carugno, R.
Onofrio, and G. Ruoso, Phys. Rev. Lett. \textbf{88}, 041804
(2002).

\bibitem{Decca}  R. S. Decca, D. L\'{o}pez, E. Fischbach, and D. E.
Krause, Phys. Rev. Lett. \textbf{91}, 050402 (2003); R. S. Decca
et. al, hep-ph/0310157 (to appear in Phys. Rev. D).

\bibitem{BS} M. Bostr\"{o}m and Bo E. Sernelius, Phys. Rev. A
\textbf{61} 046101 (2000).

\bibitem{LR00}  A. Lambrecht and S. Reynaud, Eur. Phys. J. D \textbf{8}, 309
(2000).

\bibitem{KMM}  G. L. Klimchitskaya, U. Mohideen, and V. M. Mostepanenko, Phys.
Rev. A \textbf{61}, 062107 (2000).

\bibitem{SL2}  V. B. Svetovoy, M. V. Lokhanin, Mod. Phys. Lett. A \textbf{15},
1437 (2000).

\bibitem{LP9}  E. M. Lifshitz and L. P. Pitaevskii, \textit{Statistical
Physics, Part 2 }(Pergamon Press, Oxford, 1980).

\bibitem{HB1}  \textit{Handbook of Optical Constants of Solids}, edited by
E.D. Palik (Academic Press, 1995).

\bibitem{AKB}  D. E. Aspnes, E. Kinsbron, and D. D. Bacon, Phys. Rev. B
\textbf{21}, 3290 (1980).

\bibitem{Wang}  Yu. Wang et al., Thing Solid Films, \textbf{313-314}, 232
(1998).

\bibitem{SEN}  J. Sotelo, J. Ederth, and G. Niklasson, Phys. Rev. B \textbf{%
67}, 195106 (2003).

\bibitem{KF}  K. L. Kliewer and R. Fuchs, Phys. Rev. \textbf{172}, 607 (1968).

\bibitem{VSHOVA}  L. Vazquez et al., Surf. Sci. \textbf{345},17 (1996).

\bibitem{ST}  C. C. Speake and C. Trenkel, Phys. Rev. Lett. \textbf{90},
160403 (2003).

\bibitem{FBB}  C. J. Fall, N. Binggeli, and A. Baldereschi, Phys. Rev. Lett.
\textbf{88}, 156802 (2002).

\bibitem{Esquiv}  R. Esquivel, C. Villarreal, and W. L. Moch\'{a}n, Phys. Rev.
A {\bf 68}, 052103 (2003).
\end{thebibliography}
\end{document}